\documentclass
[aps,prd,twocolumn,showpacs,nofootinbib,amsmath,amssymb,floatfix,superscriptaddress,showkeys]{revtex4-1}
\usepackage[colorlinks, linkcolor=blue, anchorcolor=blue, citecolor=blue, colorlinks=true, urlcolor=blue]{hyperref}
\usepackage{amssymb}
\usepackage{amsmath}
\usepackage{graphicx}
\usepackage{indentfirst}
\usepackage{subfigure}
\usepackage{ulem}
\usepackage{bm}
\usepackage{xspace}
\usepackage{threeparttable}
\usepackage{gensymb}
\usepackage{multirow}
\usepackage{epstopdf}
\usepackage{breakurl}
\usepackage{color}
\usepackage{multirow}
\makeatother
\allowdisplaybreaks

\newcommand{\argo}{ARGO-YBJ\xspace}
\newcommand{\fermi}{\textit{Fermi}-LAT\xspace}
\newcommand{\hess}{H.E.S.S.\xspace}

\begin{document}
\title{Limits on axion-like particles from Mrk 421 with 4.5-year period observations\\ by \argo and \fermi}
\author{Hai-Jun Li}
\affiliation{Key Laboratory of Particle Astrophysics, Institute of High Energy Physics, Chinese Academy of Sciences, Beijing 100049, China}
\affiliation{School of Physics, University of Chinese Academy of Sciences, Beijing 100049, China}
\author{Jun-Guang Guo}
\affiliation{Key Laboratory of Particle Astrophysics, Institute of High Energy Physics, Chinese Academy of Sciences, Beijing 100049, China}
\affiliation{School of Physics, University of Chinese Academy of Sciences, Beijing 100049, China}
\author{Xiao-Jun Bi}
\affiliation{Key Laboratory of Particle Astrophysics, Institute of High Energy Physics, Chinese Academy of Sciences, Beijing 100049, China}
\affiliation{School of Physics, University of Chinese Academy of Sciences, Beijing 100049, China}
\author{Su-Jie Lin}
\affiliation{School of Physics and Astronomy, Sun Yat-Sen University, Zhuhai 519082, GuangDong, China}
\affiliation{Key Laboratory of Particle Astrophysics, Institute of High Energy Physics, Chinese Academy of Sciences, Beijing 100049, China}
\author{Peng-Fei Yin}
\affiliation{Key Laboratory of Particle Astrophysics, Institute of High Energy Physics, Chinese Academy of Sciences, Beijing 100049, China}

\date{\today}

\begin{abstract}

In this work, we investigate the axion-like particle (ALP)-photon oscillation effect in the spectra of the blazar Markarian 421 (Mrk 421) using 4.5 years of the Astrophysical Radiation with Ground-based Observatory at YangBaJing (ARGO-YBJ) and \textit{Fermi} Large Area Telescope (\textit{Fermi}-LAT) data.
These data are collected during the common operation time, which cover ten activity phases of Mrk 421. No significant ALP-photon oscillation effect is confirmed. On the other hand, not all the observations of the ten phases can be individually used to set the 95\% confidence level ($\rm C.L.$) constraint on the ALP parameter space.
However, the constraint can be significantly improved if the analyses for the ten phases are combined.
We find that the upper limit at 95\% $\rm C.L.$ on the ALP-photon coupling $g_{a\gamma}$ set by the ARGO-YBJ and \textit{Fermi}-LAT observations of Mrk 421 is within $\sim [2\times 10^{-11}, \ 6\times 10^{-11}] \rm \, GeV^{-1}$ for the ALP mass of $\sim 5\times 10^{-10}$ eV $\lesssim m_a \lesssim 5\times 10^{-7}\, \rm eV$.
%\pacs{ }
%\keywords{ }

\end{abstract}
\maketitle

\section{Introduction}%%%%%%%%%%%%%%%%%%%%%%%%%%%%%%%%%%%%%%%%%Introduction

In order to solve the strong \textit{CP} problem in Quantum chromodynamics (QCD), Peccei and Quinn proposed a new $U(1)$ symmetry in 1977 \cite{Peccei:1977ur,Peccei:1977hh}.
Soon thereafter, the existence of the axion under this symmetry was recognized independently by Weinberg \cite{Weinberg:1977ma} and Wilczek \cite{Wilczek:1977pj}.
The QCD axion can be described as the pseudo Nambu-Goldstone boson of a spontaneously broken $U(1)_{\rm PQ}$ symmetry \cite{Kim:1986ax}.
Apart from the canonical QCD axion, various axion-like particles (ALPs) have also been proposed in the new physics model beyond the standard model, such as string theory \cite{Svrcek:2006yi,Arvanitaki:2009fg,Marsh:2015xka}.
The axion-like particle (ALP) mass $m_a$ and the coupling constant with photons $g_{a\gamma}$ are taken to be two independent parameters in the research.
This is different from the scenario of the QCD axion, where these two parameters are related to each other.
If ALPs are produced non-thermally in the early Universe, they may account for all or a significant fraction of the cold dark matter \cite{Preskill:1982cy, Abbott:1982af, Dine:1982ah, Khlopov:1999tm, Sikivie:2009fv}.

ALPs could be detected through their coupling to photons in the magnetic field in the laboratory \cite{Kaplan:1985dv,Sikivie:1983ip,Raffelt:1987im}, such as CAST \cite{Zioutas:1998cc,Anastassopoulos:2017ftl}, PVLAS \cite{Zavattini:2005tm,Bregant:2008yb}, OSQAR \cite{Pugnat:2007nu,Ballou:2015cka}, and ABRACADABRA \cite{Kahn:2018fgp, Ouellet:2018beu}.
The coupling between the ALP and photons would also lead to ALP-photon oscillation for the photons ejecting from the high energy $\gamma$-ray sources which are far from the Earth and would modify their $\gamma$-ray energy spectra \cite{Raffelt:1987im}.
This effect could lead to the observable modifications in the $\gamma$-ray telescopes \cite{DeAngelis:2007dqd,Hooper:2007bq}.
Many works have been performed to study this effect for many astrophysical sources and set constraints on the ALP parameter space \cite{DeAngelis:2007dqd,Hooper:2007bq,Simet:2007sa, Belikov:2010ma, DeAngelis:2011id, Tavecchio:2012um, Tavecchio:2014yoa, Meyer:2013pny, Meyer:2014epa, Galanti:2015rda, Meyer:2016wrm, Majumdar:2017vcx, Galanti:2018nvl, Galanti:2018myb, Galanti:2018upl, Berenji:2016jji, Reesman:2014ova, Zhang:2018wpc, Liang:2018mqm, Long:2019nrz, Kohri:2017ljt, Libanov:2019fzq, Bi:2020ths, Abramowski:2013oea, TheFermi-LAT:2016zue, Guo:2020kiq, Pallathadka:2020vwu}.

In this work, we focus on the very high energy (VHE) $\gamma$-ray observations of the blazar Markarian 421 (Mrk 421), which is one of the most widely studied and brightest sources in the extragalactic TeV sky with the redshift of $z_0$ = 0.031.
It was firstly detected at VHE by the Whipple Observatory in 1992 \cite{Punch:1992xw} and has been well detected with many imaging atmospheric Cherenkov telescopes \cite{Aharonian:2002fk, Aharonian:2005ib, Albert:2006jd, Acciari:2009sta, Abdo:2011zz,collaboration:2011esa,Bartoli:2015cvo,Acciari:2019zgl}.
Mrk 421 is classified as a high synchrotron-peaked BL Lac object and is a very active blazar with major outbursts \cite{Abdo:2009iq}, which are composed of many short flares in both the X-ray and $\gamma$-ray regions.

Thanks to the results of Astrophysical Radiation with Ground-based Observatory at YangBaJing (\argo) \cite{collaboration:2011esa,Bartoli:2015cvo} and \textit{Fermi} Large Area Telescope (\fermi) \cite{Abdo:2011zz}, the high energy component of the spectral energy distribution of Mrk 421 has been completely covered in the $\gamma$-ray band from 0.1 GeV to 10 TeV.
Ref.~\cite{Bartoli:2015cvo} reported the 4.5-year multi-wavelength data recorded from August 2008 to February 2013.
This period contains the ten steady and flaring phases of Mrk 421.
In this work, we consider the effect of the ALP-photon oscillation in the high energy $\gamma$-ray spectra of Mrk 421 with these ten phases and combine the results of all the phases together to set constraint on the ALP parameter space.

This paper is organized as follows.
In Section \ref{section_alp-p}, we introduce the propagation of high energy $\gamma$-ray with the ALP-photon oscillation in three parts, including the source region, the extragalactic space, and the Milky Way region.
In Section \ref{section_data}, we describe the observation data of Mrk 421 by \argo and \fermi
in the common operation time used in this work.
In Section \ref{section_met}, we introduce the data fitting and statistical methods.
In Section \ref{section_res}, we give the constraints on the ALP parameter space from the Mrk 421 observations.
The conclusion is given in Section \ref{section_sum}.

\section{Propagation of the ALP-photon system}%%%%%%%%%%%%%%%%%%%%%%%%%%%%%%%%%%%%%%%%%Propagation
\label{section_alp-p}

The ALP-photon oscillation effect occurring in the magnetic field would modify the $\gamma$-ray energy spectra of astrophysical sources, which are far from the Earth \cite{DeAngelis:2007dqd, Hooper:2007bq}.
The ALP-photon coupling is described by the following Lagrangian \cite{Raffelt:1987im}
\begin{eqnarray}
\mathcal{L}_{a\gamma}=-\frac{1}{4}g_{a\gamma}aF_{\mu\nu}\tilde{F}^{\mu\nu}=g_{a\gamma}a\textbf{E}\cdot\textbf{B},
\end{eqnarray}
where $g_{a\gamma}$ is the coupling constant, $a$ is the ALP field, $F_{\mu\nu}$ is the electromagnetic field tensor, $\tilde{F}^{\mu\nu}$ is the dual tensor, and $\textbf{E}$ and $\textbf{B}$ are the electric and magnetic fields, respectively.
The ALP and photon are interconvertible in the external magnetic field
and could be described by the ALP-photon beam \cite{DeAngelis:2011id}
\begin{eqnarray}
\Psi = \left(A_1, A_2, a\right)^T,
\end{eqnarray}
where $A_1$ and $A_2$ represent the photon transverse polarization states in the directions of $x_1$ and $x_2$ which are perpendicular to the propagation direction $x_3$, respectively.

The density matrix of the ALP-photon system is given by
\begin{eqnarray}
\rho=\Psi \otimes \Psi^\dagger.
\end{eqnarray}
After oscillation in numerous consecutive domains, the final density matrix of the ALP-photon system is
\begin{eqnarray}
\rho\left(s\right)=T\left(s\right)\rho\left(0\right)T^\dagger\left(s\right),
\end{eqnarray}
where $T\left(s\right)$ is the whole transfer matrix for the propagation distance $s$. The initial beam state $\rho(0)$ is assumed to be
\begin{eqnarray}
\rho(0)=\frac{1}{2}{\rm diag}\left(1, 1, 0\right).
\end{eqnarray}
The final survival probability of the photon in the ALP-photon system is given by \cite{DeAngelis:2011id}
\begin{eqnarray}
P_{\gamma\gamma}={\rm Tr}\left(\left(\rho_{11}+\rho_{22}\right)T\left(s\right)\rho\left(0\right)T^\dagger\left(s\right)\right),\label{pp}
\end{eqnarray}
with
\begin{eqnarray}
\rho\left(s\right)_{11}={\rm diag}\left(1,0,0\right),\quad\rho\left(s\right)_{22}={\rm diag}\left(0,1,0\right).
\end{eqnarray}

The ALP-photon conversion would become maximal and energy-independent in the strong mixing regime $E_{\rm crit} \lesssim E \lesssim E_{\rm max}$ with \cite{Meyer:2014epa, Galanti:2018nvl}
\begin{eqnarray}
\begin{aligned}
E_{\rm crit} &= \frac {{| m_a^2 - \omega_{\rm pl}^2|}}{2g_{a\gamma}B}\\
&\simeq 2.5 \ {\rm GeV} \ |m_{\rm neV}^2-1.4\times 10^{-3} \ n_{{\rm cm}^{-3}}|g_{11}^{-1}B_{\mu \rm G}^{-1},
\end{aligned}
\end{eqnarray}
and
\begin{eqnarray}
\begin{aligned}
E_{\rm max} &= \frac{90\pi}{7\alpha}\frac{B_{\rm cr}^2g_{a\gamma}}{B} \simeq 2.12 \times 10^6 \ {\rm GeV} \ g_{11}B_{\rm \mu G}^{-1},
\end{aligned}
\end{eqnarray}
where $m_a$ is the ALP mass, $\omega_{\rm pl}\sim \sqrt{4\pi \alpha n_e/m_e}$ is the plasma frequency, $B$ is the external magnetic field, $n_e$ is the electron density, $\alpha$ is the fine-structure constant, and $B_{\rm cr}=m^2_e/|e| \sim 4.41 \times 10^{13} \ \rm{G}$ is the critical magnetic field. In the above equations, we have used the notations $m_{\rm neV} \equiv m_a/1 \ \rm {neV} $, $n_{{\rm cm}^{-3}} \equiv n_e/1 \ \rm {cm}^{-3} $, $g_{11}\equiv g_{a\gamma}/10^{-11} \ \rm{GeV}^{-1}$, and  $B_{\rm \mu G}\equiv B /1  \ \mu\rm G$.

In order to obtain the transfer matrix for the ALP-photon system, the propagation process of the system is divided into three parts, including the propagations in the source region, the extragalactic space, and the Milky Way region \cite{DeAngelis:2007dqd, Hooper:2007bq}.

%%%%%%%%%%%%%%%%%%%%%%%%%%%%%%%%%%%%%%%%%%%%%%%%%%%%%%TAB.1
\begin{table}[htb]
\caption{The benchmark values of the parameters for the BJMF model in the ten phases of Mrk 421.}
\begin{ruledtabular}
\begin{tabular}{lccc}
Phase &$B_0$(G)   & $ \delta_{\rm D} $    & $n_0$($10^{3} \ \rm cm^{-3}$)     \\
\hline
S1   & 0.048  &  38  & 1.919 \\
S2   & 0.17  &  15  & 3.009 \\
OB   & 0.054  &  35  & 2.228 \\
F1   & 0.14  &  10  & 0.269  \\
F2   & 0.092  &  17  & 0.825  \\
F3   & 0.080  &  41  & 1.544 \\
F4   & 0.033  &  35  & 3.527 \\
F5   & 0.072  &  31  & 2.296 \\
F6   & 0.085  &  15  & 8.809 \\
F7   & 0.115  &  30  & 1.963 \\
\end{tabular}
\end{ruledtabular}
\label{tab_1}
\end{table}
%%%%%%%%%%%%%%%%%%%%%%%%%%%%%%%%%%%%%%%%%%%%%%%%%%%%%%%%%

Firstly, we neglect the internal $\gamma$-ray absorption and the ALP-photon oscillation within the broad line region of the source, and only consider the ALP-photon oscillation effect in the blazar jet magnetic field (BJMF).
Following Refs.~\cite{Tavecchio:2014yoa, Galanti:2018upl}, here the BJMF of Mrk 421 is considered as the BL Lac type with the transverse magnetic field $B_{\rm jet}(r)$ and the electron density $n_{\rm el}(r)$ profiles.
The radial profile of the magnetic field strength is \cite{Begelman:1984mw, Ghisellini:2009wa, Pudritz:2012xj}
\begin{eqnarray}
B_{\rm jet}\left(r\right)=B_0\left(\frac{r}{r_{\rm VHE}}\right)^{-1},
\label{br}
\end{eqnarray}
where $r_{\rm VHE}$ is the distance of the VHE emission site to the central black hole and $B_0$ is the magnetic field strength at $r_{\rm VHE}$.
The modified model for the electron density distribution is \cite{OSullivan:2009dsx}
\begin{eqnarray}
n_{\rm el}\left(r\right)=n_0\left(\frac{r}{r_{\rm VHE}}\right)^{-2},
\end{eqnarray}
where $n_0$ is the electron density at $r_{\rm VHE}$.
These equations hold in the co-moving frame of the blazar jet.
The photon energy $E_j$ in this frame is related to the energy $E_L$ in the laboratory frame with the transformation $E_j= E_L/\delta_{\rm D}$, where $\delta_{\rm D}$ is the Doppler factor.
In the region with $r > 1$ kpc, the strength of BJMF is set to be zero.
More details about this BJMF model can be found in Refs.~\cite{Tavecchio:2014yoa, Galanti:2018upl}.

In principle, the parameters for the BJMF model can be derived from the fit to the data using the synchrotron self-Compton model.
The best-fit values of $B_0$ and $\delta_{\rm D}$ in the ten phase of Mrk 421 given by Ref.~\cite{Bartoli:2015cvo} are listed in Table~\ref{tab_1}.
The electron distribution within the emitting plasma is assumed to be a broken power law in Ref.~\cite{Bartoli:2015cvo}.
Using the best-fit values of the energy normalization $u_e$ and the first index of the broken power law $p_1$, we derive the benchmark values of $n_0$ listed in Table~\ref{tab_1}.

It is difficult to derive the precise value of $r_{\rm VHE}$ from the observations.
$r_{\rm VHE}$ can be roughly estimated as $r_{\rm VHE}\sim R_{\rm VHE}/\theta_{\rm jet}$, where $R_{\rm VHE}$ is the radius of the VHE emitting plasma blob and $\theta_{\rm jet}$ is the angle between the jet axis and the line of sight. In the analysis of ARGO-YBJ $R_{\rm VHE}$ for all the phases are arbitrarily set to be $10^{16}$ cm, while no $\theta_{\rm jet}$ is provided in Ref.~\cite{Bartoli:2015cvo}.
The fit using a multiple-flare model leads to $\theta_{\rm jet}\in [0.38^\circ-1.8^\circ]$ in Ref.~\cite{Hervet:2019caf}, which is consistent with $\theta_{\rm jet}=1.3^\circ $ derived in Ref.~\cite{Banerjee:2019oyp}.
In Ref.~\cite{Celotti:2007rb}, the values of $R_{\rm VHE}$ and $\theta_{\rm jet}$ of Mrk 421 are found to be $6\times 10^{15}$cm and $2.0\degree$, respectively.
Using these values, we find $r_{\rm VHE}\sim \mathcal{O}(10^{17})-\mathcal{O}(10^{18})$ cm. In the following analysis, we shall take $r_{\rm VHE}=10^{17}$ cm as the benchmark parameter for all the phases.

After leaving the jet, the ALP-photon system enters the host galaxy.
Following Refs.~\cite{Tavecchio:2012um, Galanti:2018upl}, we do not consider the ALP-photon oscillation effect in the magnetic field of the host galaxy since it is too small.
In general, the blazar may be located at a rich cluster, where the turbulent inter-cluster magnetic field is about $\mathcal{O}(1)$ $\mu$G \cite{Carilli:2001hj, Govoni:2004as, Subramanian:2005hf}.
We find that the ALP-photon oscillation could be significant in this magnetic field.
However, there is no evidence that Mrk 421 is reside in such a rich environment. Therefore, we do not consider the
ALP-photon oscillation effect in the inter-cluster magnetic field.

\begin{figure}%%%%%%%%%%%%%%%%%%%%%%%%%%%%%%%%%%%%%%%%%FIG.fran
\centering
\includegraphics[width=0.5\textwidth]{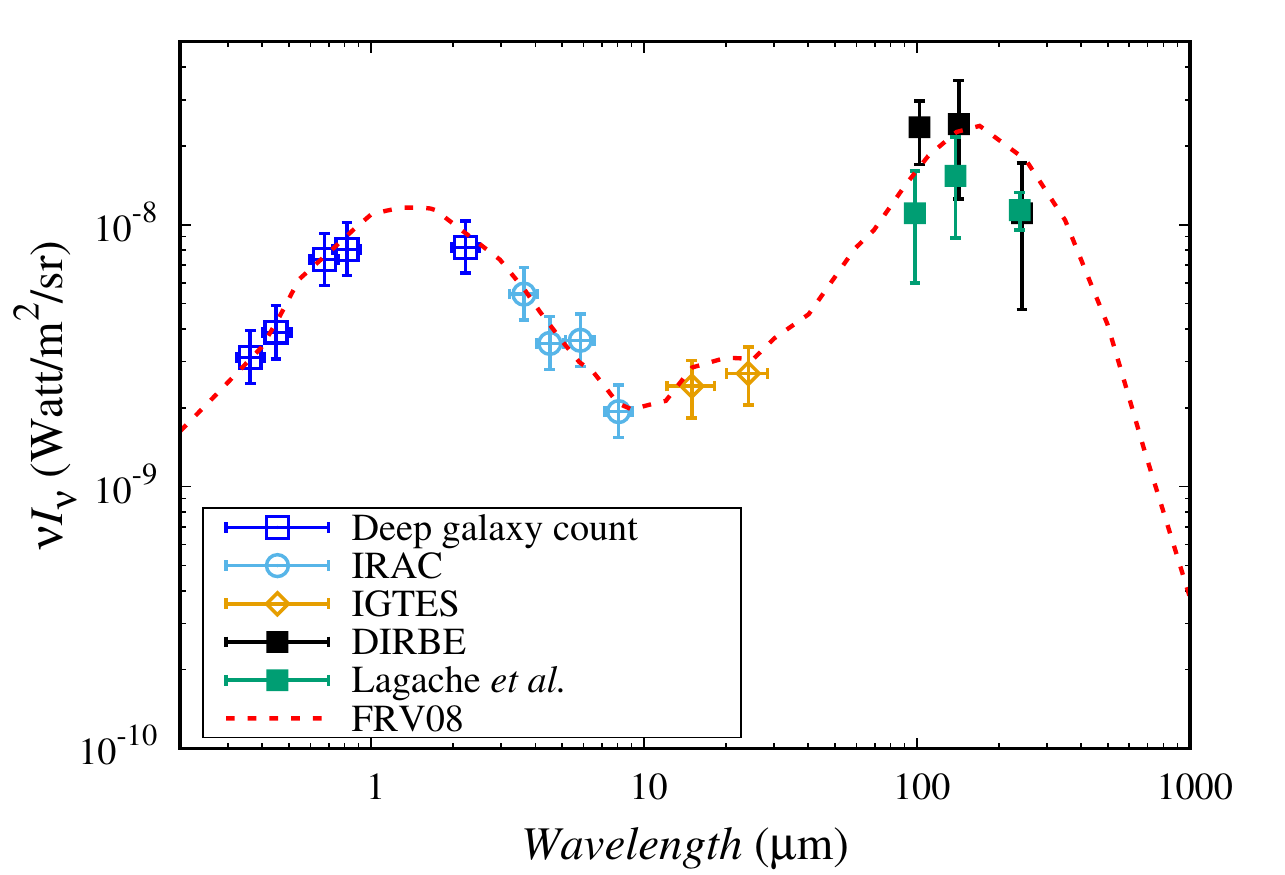}
\caption{The cosmic infrared background spectrum (dashed line) from the multi-wavelength reference model with the galaxy evolution effect named FRV08 model in Ref.~\cite{Franceschini:2008tp}.
The observed data from Refs.~\cite{Madau:1999yh, Fazio:2004mu, Elbaz:2002vd, Papovich:2004vh, Madau:1999yh, Lagache:1999ji} are also shown.}
\label{fig_fran}
\end{figure}

The upper limit of the extragalactic magnetic field on the largest cosmological scale is $ \mathcal{O}$(1) nG \cite{Pshirkov:2015tua}.
Its actual value is not clear and would be much lower than this upper limit \cite{Ade:2015cva, Zhang:2018wpc}.
In this work, we do not consider the effect of this magnetic field for the ALP-photon system propagation in the extragalactic space and only focus on the attenuation effect induced by the extragalactic background light (EBL) due to the pair production process
$\gamma + \gamma_{\rm BG} \to e^+ + e^-$.
This attenuation effect can be characterized by the factor of $e^{-\tau}$ with the optical depth \cite{Belikov:2010ma, Franceschini:2008tp}
\begin{eqnarray}
\tau=c \int_0^{z_0} \frac{{\rm d}z}{(1+z)H(z)}\int_{E_{\rm th}}^{\infty}{\rm d}\omega\frac{{\rm d}n(z)}{{\rm d}\omega}\bar{\sigma}\left(E_\gamma,\omega,z\right),~
\end{eqnarray}
where $z_0$ is the redshift of the source, $H(z)$ is the rate of the Hubble expansion, $E_{\rm th}$ is the threshold energy, $\bar{\sigma}$ is the integral cross section of the pair production, $E_\gamma$ and $\omega$ are the source and background photon energies, respectively, and ${\rm d}n/{\rm d}\omega$ is the proper number density of the EBL.
The EBL model used here is taken from Ref.~\cite{Franceschini:2008tp} and is shown in Fig.~\ref{fig_fran}.

For the propagation of the ALP-photon system in the Milky Way region, we consider the ALP-photon oscillation effect in the Galactic magnetic field. Here we neglect the random component on the small scale and take the regular component of the Galactic magnetic field on the large scale from Ref.~\cite{Jansson:2012rt}.

\section{Gamma-ray data of Mrk 421 from \argo and \fermi}%%%%%%%%%%%%%%%%%%%%%%%%%%%%%%%%%%%%%%%%%Data
\label{section_data}

The \argo \cite{Bartoli:2011qe} detector, located at Yangbajing Cosmic Ray Observatory (Tibet, P.R. China, $90.5\degree$ East, $30.1\degree$ North), is an air shower array exploiting the full coverage approach at very high altitude, with the aim of studying the VHE $\gamma$-ray astronomy and cosmic-ray physics.

In Ref.~\cite{Bartoli:2015cvo}, the \argo collaboration reported the spectral variation of Mrk 421 at different wavebands and divided the whole observation period into ten phases according to the large X-ray and GeV $\gamma$-ray flares. The $\gamma$-ray spectra of Mrk 421 at lower energies in the common operation time from \fermi \cite{Atwood:2009ez} are also analysed in Ref.~\cite{Bartoli:2015cvo}.
The analysis is performed with the standard ScienceTool and the corresponding threads provided by \fermi.

The ten phases of the \argo observation for Mrk 421 \cite{Bartoli:2015cvo} are described as follows.
Mrk 421 showed a low activity at all wavebands from August 2008 to June 2009. This is marked as Steady 1 (S1) phase.
Then the source entered a long lasted outburst phase from June 2009 to June 2010, which is denoted as Outburst (OB) phase.
During this active phase, three large flares Flare 1 (F1), Flare 2 (F2), and Flare 3 (F3) were clearly detected.
After May 2010, Mrk 421 entered a low steady phase until October 6. This phase lasted about one month and is marked as Flare 4 (F4) phase.

Then Mrk 421 came to a long and steady phase (S2) from November 2010 to June 2012, which lasted about 1.6 years.
The embedded strong flare denoted as Flare 5 (F5) phase, which occurred in September 2011 and lasted 7 days, has been separated from the phase S2.
In the whole year of 2012, the flux of $\gamma$-ray measured by \argo \cite{2012ATel.4272} and \fermi \cite{2012ATel.4261} reached a high level from 2012 July 9 to September 17. Two peaks are marked as Flare 6 (F6, from 2012 July 9 to 21) phase and Flare 7 (F7, from 2012 July 22 to September 16) phase.

\section{Data fitting and statistical methods}%%%%%%%%%%%%%%%%%%%%%%%%%%Methods
\label{section_met}

In order to fit the experimental data of \argo and \fermi under the null hypothesis, the form of the intrinsic energy spectrum of Mrk 421 is taken to be a super-exponential cut-off power law (SEPWL)
\begin{eqnarray}
\Phi_{\rm int} \left( E \right) = F_0\left(\frac{E}{E_0}\right)^{-\Gamma}\exp{\left(-\left(\frac{E}{E_c}\right)^d\right)},
\end{eqnarray}
where $E_0$ is taken to be 1 GeV, $F_0$, $\Gamma$, $E_c$, and $d$ are treated as free parameters.
We also test other forms of the intrinsic spectrum including the exponential cut-off power law and the logarithmic parabola function. We find that the best-fit $\chi ^2$ under the null hypothesis of the SEPWL is the smallest.

Considering the modification of the energy spectrum induced by the ALP, the expected $\gamma$-ray energy spectrum under the ALP hypothesis is determined by the survival probability of the photon $P_{\gamma\gamma}$ in Eq.~(\ref{pp})
\begin{eqnarray}
\Phi_{\rm w \; ALP} \left( E \right) = P_{\gamma\gamma} \Phi_{\rm int} \left( E \right),
\end{eqnarray}
with the intrinsic energy spectrum $\Phi_{\rm int}(E)$.

We also take into account the energy resolution of the experiments in the analysis.
The energy resolutions of \argo and \fermi are adopted to be 13\% \cite{Bartoli:2013qxm} and 15\% \footnote{\url{https://fermi.gsfc.nasa.gov/ssc/data/analysis/documentation/Cicerone/Cicerone_Introduction/LAT_overview.html}}, respectively.
Considering the energy dispersion function $D(E',E_1,E_2)$ where $E'$ is the actual energy, the expected $\gamma$-ray flux at the detector in the energy bin between $E_1$ and $E_2$ can be derived as \cite{Guo:2020kiq}
\begin{eqnarray}
\Phi'=\frac{\int_0^{\infty}D(E',E_1,E_2)\Phi \left( E' \right){\rm d}E'}{E_2 - E_1},
\end{eqnarray}
where $\Phi(E')$ is the $\gamma$-ray spectrum before detection.
The $\chi^2$ value is given by
\begin{eqnarray}
\chi^2 = \sum_{i=1}^{N} \frac{( \Phi'_{i} - \tilde{\phi}_i )^2}{\delta_i^2},
\end{eqnarray}
where $N$ is the point number, $\Phi'_{i}$ is the expected $\gamma$-ray flux, $\tilde{\phi}_i$ is the observed flux, and $\delta_i$ is the corresponding uncertainty of the observation.

In order to set the constraint on the ALP parameter space, we define the threshold value $\chi_{\rm th}^2$ as
\begin{eqnarray}
\chi_{\rm th}^2 = \chi_{\rm min}^2 + \Delta{\chi}^2,
\end{eqnarray}
with the minimum best-fit ${\chi}_{\rm min}^2$ in the $m_a-g_{a\gamma}$ plane and the $\Delta{\chi}^2$ corresponding to the particular confidence level ($\rm C.L.$).
Due to the non-linear dependence of the spectral irregularities on the ALP parameters, we derive the value of $\Delta{\chi}^2$ from the Monte Carlo simulation \cite{TheFermi-LAT:2016zue}.

We generate 400 sets of the observed $\gamma$-ray spectra of Mrk 421 in the pseudo-experiments that are realized by Gaussian samplings \cite{Liang:2018mqm}.
For each set of the simulated spectrum, we can derive the best-fit $\chi^2$ for both the null hypothesis ${\widehat{\chi}_{\rm null}}^2$ and the ALP hypothesis ${\widehat{\chi}_{\rm w \; ALP}}^2$.
For each Monte Carlo data set, we have the test statistic (TS) value
\begin{eqnarray}
{\rm TS} ={\widehat{\chi}_{\rm null}}^2 - {\widehat{\chi}_{\rm w \; ALP}}^2.
\end{eqnarray}
Then we obtain the TS distribution under the null hypothesis for all data sets that obeys the non-central $\chi^2$ distribution.
The $\Delta{\chi}^2$ corresponding to the certain confidence level can be derived from the TS distribution with the effective degree of freedom ($\rm d.o.f.$) and the non-centrality $\lambda$. Finally, we assume that the probability distribution under the alternative hypothesis with ALP is approximated with the distribution under the null hypothesis and use the value of $\Delta{\chi}^2$ derived above to set the constraint on the ALP parameter space \cite{TheFermi-LAT:2016zue}.

%%%%%%%%%%%%%%%%%%%%%%%%%%%%%%%%%%%%%%%%%%%%%%%%%%%%%%TAB.2
\begin{table}[htb]
\caption{The best-fit values of $\chi_{\rm w/o \; ALP}^2$ in the ten phases under the null hypothesis.}
\begin{ruledtabular}
\begin{tabular}{lcr}
Phase &data point number  &   $ \chi_{\rm w/o \; ALP}^2 $         \\
\hline
S1 &16    & 18.88        \\
S2 &16    & 9.47          \\
OB &16    & 18.38          \\
F1 &7     & 1.45         \\
F2 &9     & 0.99            \\
F3 &9     & 6.61             \\
F4 &11    & 3.91           \\
F5 &7     & 7.18         \\
F6 &11    & 7.88             \\
F7 &15    & 5.20           \\
\end{tabular}
\end{ruledtabular}
\label{tab_2}
\end{table}
%%%%%%%%%%%%%%%%%%%%%%%%%%%%%%%%%%%%%%%%%%%%%%%%%%%%%%%%%

%%%%%%%%%%%%%%%%%%%%%%%%%%%%%%%%%%%%%%%%%%%%%%%%%TAB.3
\begin{table}[!htbp]
\caption{The minimum best-fit values of ${\chi}_{\rm min}^2$ in the ten phases under the ALP hypothesis. The effective $\rm d.o.f.$ of the TS distributions and the values of $\Delta{\chi}^2$ corresponding to 95\% $\rm C.L.$ are also listed. The combined results for the ten phases are also shown. The values of the BJMF parameters are given by Table~\ref{tab_1}.}
\begin{ruledtabular}
\begin{tabular}{lccr}
%\hline
%\hline
Phase &${\chi}_{\rm min}^2$   & effective $\rm d.o.f.$  & $\Delta{\chi}^2$ \\
\hline
S1    &   11.81    &   4.83  & 10.83      \\
S2    &   6.49     &   4.86  & 11.88         \\
OB    &   12.85    &   4.98  & 11.06        \\
F1    &   0.77     &   1.59  & 5.20      \\
F2    &   0.49     &   3.03  & 7.89         \\
F3    &   2.94     &   3.03  & 7.89          \\
F4    &   2.27     &   1.16  & 4.26        \\
F5    &   5.45     &   1.17  & 4.28         \\
F6    &   4.79     &   1.16  & 4.26          \\
F7    &   2.57     &   4.76  & 10.72         \\
\hline
combined   &   72.39  & 7.09  & 14.22    \\
%\hline
%\hline
\end{tabular}
\end{ruledtabular}
\label{tab_3}
\end{table}
%%%%%%%%%%%%%%%%%%%%%%%%%%%%%%%%%%%%%%%%%%%%%%%%%%%

\begin{figure*}%%%%%%%%%%%%%%%%%%%%%%%%%%%%%%%%%%%%%%%FIG.dnde
\includegraphics[width=1\textwidth]{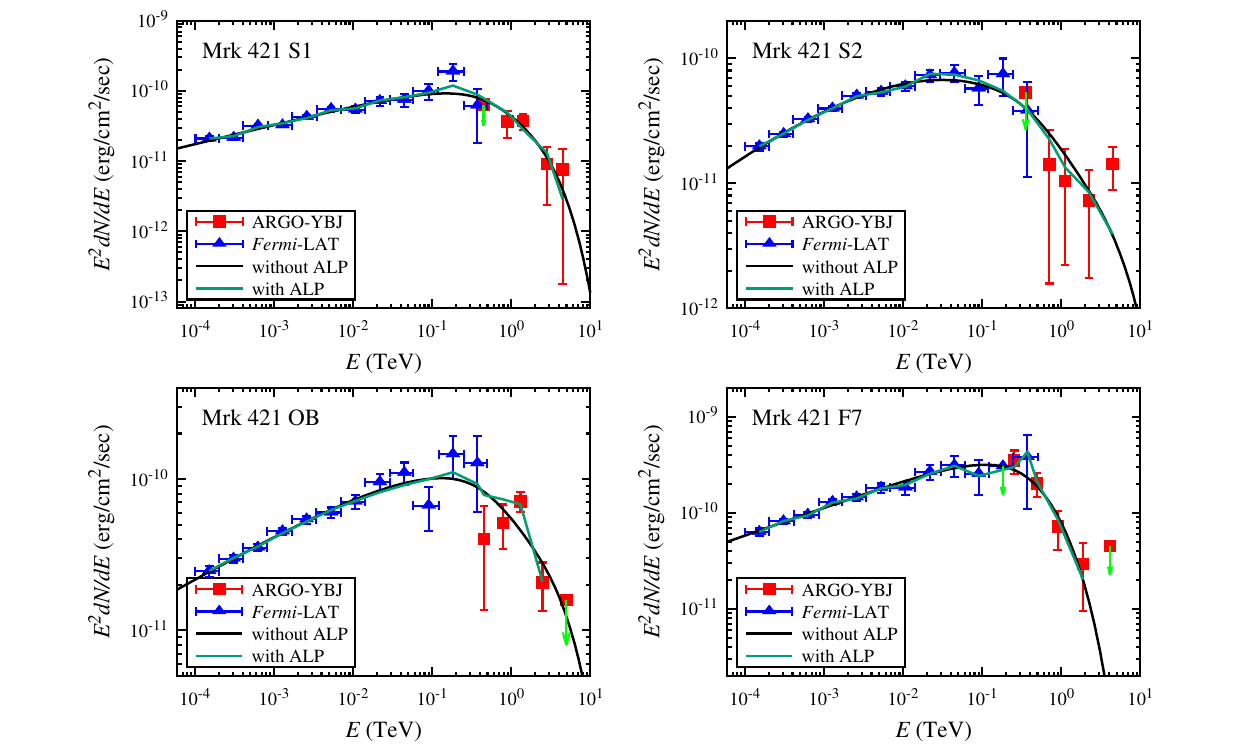}
\caption{The best-fit $\gamma$-ray spectra of Mrk 421 in the phases S1 (top left), S2 (top right), OB (bottom left), and F7 (bottom right). The black lines represent the spectra under the null hypothesis with $\chi_{\rm w/o \; ALP}^2$ = 18.88, 9.47, 18.38, and 5.20 in the four phases. The green lines represent the spectra under the ALP hypothesis with ${\chi}_{\rm min}^2$ = 11.81, 6.49, 12.85, and 2.57 in the four phases. The parameters for  the BJMF model are given by Table~\ref{tab_1}.
The experimental data are from \argo and \fermi \cite{Bartoli:2015cvo}.}
\label{fig_dnde}
\end{figure*}

\begin{figure*}[!htbp]%%%%%%%%%%%%%%%%%%%%%%%%%%%%%%%%%%%%%%%%%FIG.con1
\centering
\includegraphics[width=1\textwidth]{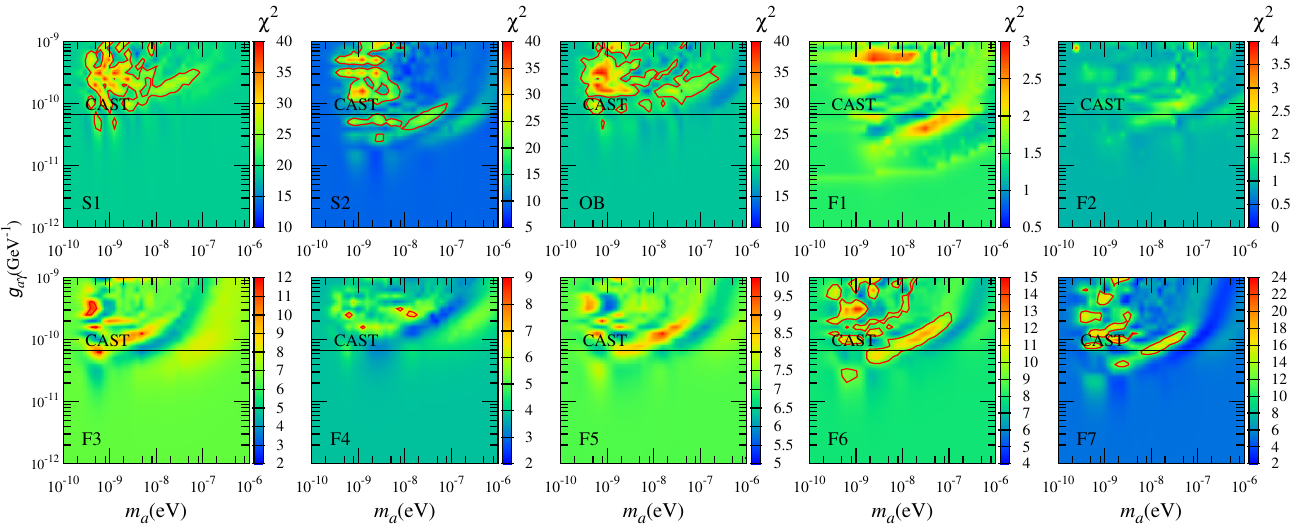}
\caption{The distributions of $\chi_{\rm w \; ALP}^2$ in the $m_a-g_{a\gamma}$ plane for the ten phases of Mrk 421. The parameters for the BJMF model are given by Table~\ref{tab_1}.
The red contours represent the excluded regions at 95\% $\rm C.L.$ in the phases S1, S2, OB, F3, F4, F6, and F7. The horizontal line represents the upper bound on $g_{a\gamma}$ set by CAST \cite{Anastassopoulos:2017ftl} of $g_{a\gamma} < 6.6\times 10^{-11}$ $\rm GeV^{-1}$.
}
\label{fig_chi2_1}
\end{figure*}

\begin{figure*}[!h]%%%%%%%%%%%%%%%%%%%%%%%%%%%%%%%%%%%%%%%%%FIG.ts
\centering
\includegraphics[width=0.95\textwidth]{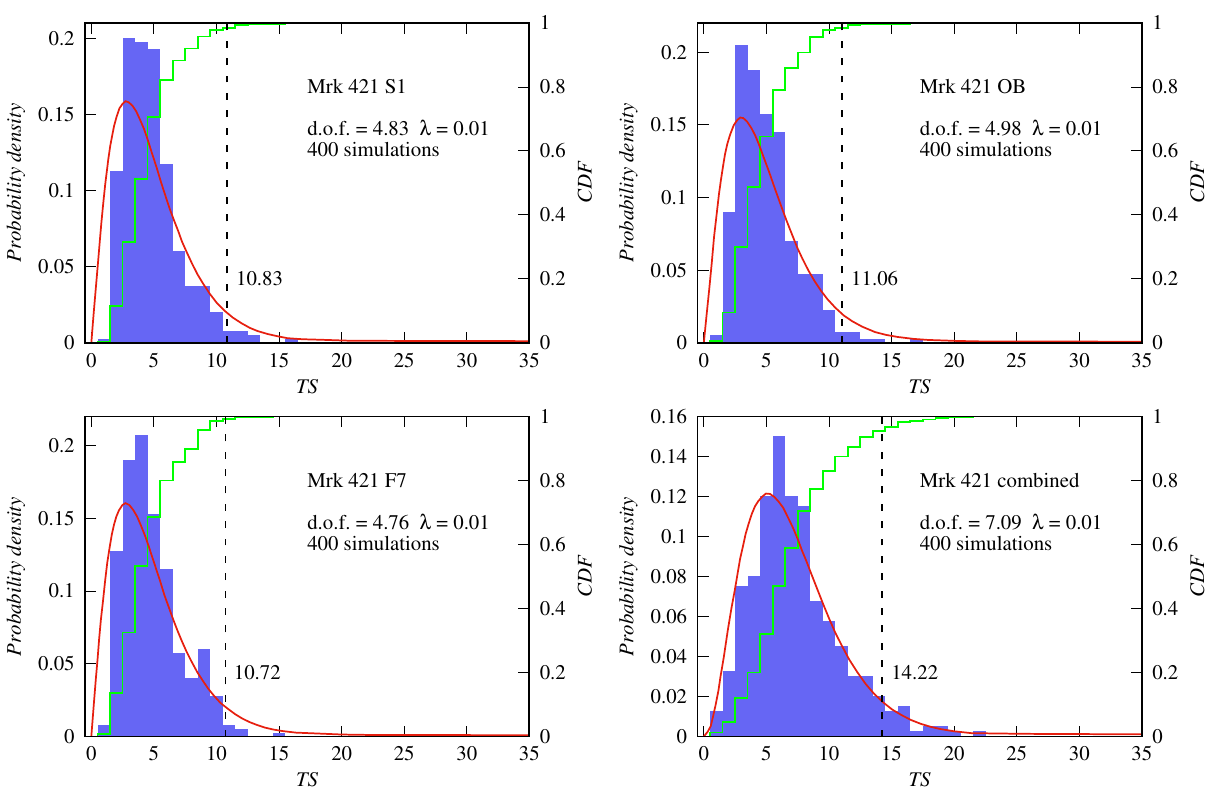}
\caption{The TS distributions of the phase S1 (top left), OB (top right), F7 (bottom left), and the combined phases (bottom right) with the BJMF parameters given by Table~\ref{tab_1}. The red lines represent the fitted non-central $\chi^2$ distributions. The green lines represent the CDF of the TS distributions.}
\label{fig_ts_s1_s2}
\end{figure*}

\begin{figure*}[!htbp]%%%%%%%%%%%%%%%%%%%%%%%%%%%%%%%%%%%%%%%%%FIG.con3
\centering
\includegraphics[width=0.55\textwidth]{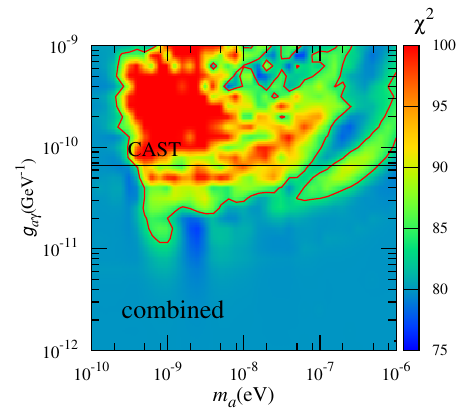}
\caption{The $\chi_{\rm w \; ALP}^2$ distribution of the combined phases with the BJMF parameters given by Table~\ref{tab_1}. The red contour represents the excluded region at 95\% $\rm C.L.$
}
\label{fig_chi2_3}
\end{figure*}

\section{Results}%%%%%%%%%%%%%%%%%%%%%%%%%%%%%%%%%%%%%%%%%Results
\label{section_res}

In this section, we set constraints on the ALP parameter space using the data of \argo and \fermi in the ten phases of Mrk 421.
The best-fit values of ${\chi}_{\rm w/o \; ALP}^2$ under the null hypothesis are listed in Table~\ref{tab_2}.
The reduced $\chi^2$ under the null hypothesis in the ten phases are around the average value 1.04.
Only the reduced $\chi^2$ in the phase F5 is large as 2.39.
In Fig.~\ref{fig_dnde}, we also give the best-fit $\gamma$-ray spectra for the phases of S1, S2, OB, and F7 under the null and ALP hypotheses.
We can see that the null hypothesis can well fit the \argo and \fermi data.

The distributions of $\chi_{\rm w \; ALP}^2$ under the ALP hypothesis in the ten phases are shown in Fig.~\ref{fig_chi2_1} with the benchmark values of the BJMF parameters.
The minimum best-fit values of ${\chi}_{\rm min}^2$ in the $m_a-g_{a\gamma}$ plane for the ten phases are listed in Table~\ref{tab_3}.

Then we can derive the TS distributions for the ten phases. We find that the non-centralities of all the TS distributions are about 0.01.
The effective $\rm d.o.f.$ of the distributions and the threshold values of $\Delta{\chi}^2$ corresponding to $95\%$ $\rm C.L.$ are listed in
Table~\ref{tab_3}. In Fig.~\ref{fig_ts_s1_s2}, we plot the TS distributions for the phases S1, OB, and F7. The red lines represent the fitted non-central $\chi^2$ distributions with the effective $\rm d.o.f.$ of 4.83, 4.98, and 4.76 for the phases S1, OB, and F7, respectively.
The green lines represent the cumulative distribution functions (CDF) of the TS distributions. Using these functions, we can derive the values of $\Delta{\chi}^2$ corresponding to the 95\% $\rm C.L.$ threshold as 10.83, 11.06, and 10.72 for the phases S1, OB, and F7, respectively.

\begin{figure}%%%%%%%%%%%%%%%%%%%%%%%%%%%%%%%%%%%%%%%%%FIG.con
\centering
\includegraphics[width=0.5\textwidth]{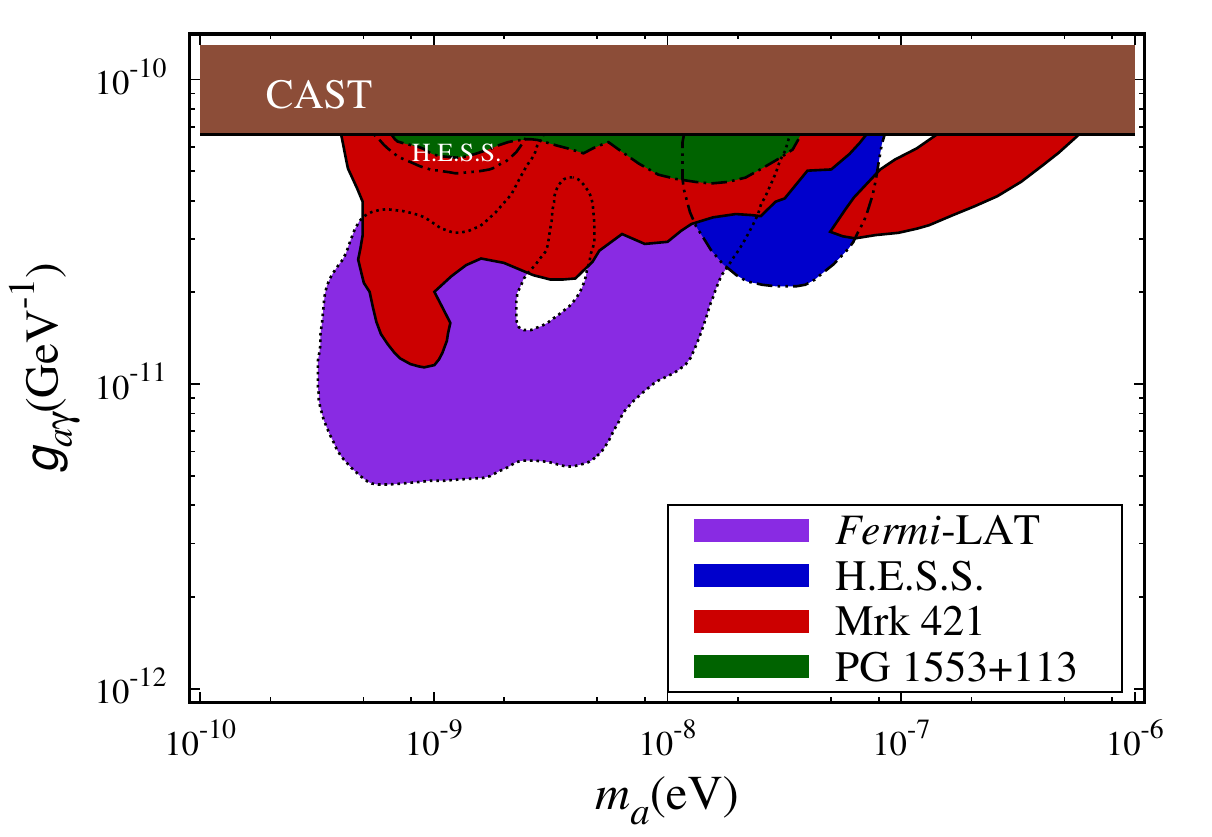}
\caption{The excluded regions at 95\% $\rm C.L.$ set by the Mrk 421 observations of \argo and \fermi. The red contour represents the result derived with the BJMF parameters given by Table~\ref{tab_1}. For comparison, the constraints set by CAST \cite{Anastassopoulos:2017ftl}, the PKS 2155-304 observation of \hess \cite{Abramowski:2013oea}, and the NGC 1275 observation of \fermi \cite{TheFermi-LAT:2016zue} are shown. We also show the constraint set by the PG 1553+113 observations using the data from $\rm \hess ~II$ and \fermi \cite{Guo:2020kiq}.}
\label{fig_contour}
\end{figure}

With the values of $\Delta{\chi}^2$ for all the phases, the constraints on the ALP parameter space at 95\% $\rm C.L.$ are shown in Fig.~\ref{fig_chi2_1}.
It can be seen that the data sets of the ten phases set very different constraints.
Using the data of the phases S1, S2, OB, F6, and F7, we could find some explicit excluded parameter regions at 95\% $\rm C.L.$
For the other phases, the 95\% $\rm C.L.$ constraint cannot be significantly set in the $m_a-g_{a\gamma}$ plane. This is because that the observations in these phases provide too few data points.

In order to make a reliable implication, we combined the results in the ten phases together.
The TS distribution and the combined $\chi_{\rm w \; ALP}^2$ distribution in the $m_a-g_{a\gamma}$ plane are shown in Fig.~\ref{fig_ts_s1_s2} and \ref{fig_chi2_3}, respectively. The red contour in Fig.~\ref{fig_chi2_3} represents the upper limit at 95\% $\rm C.L.$ Compared with the limit set by CAST \cite{Anastassopoulos:2017ftl} about $g_{a\gamma} < 6.6\times 10^{-11}$ $\rm GeV^{-1}$, the upper limit on the ALP-photon coupling $g_{a\gamma}$ set by the \argo and \fermi observations of Mrk 421 is within  $\sim [2 \times 10^{-11}, \ 6\times 10^{-11}] \rm \, GeV^{-1}$ for the ALP mass of $\sim 5\times 10^{-10}$ eV $\lesssim m_a \lesssim 5\times 10^{-7}$ eV.

For comparison, we also show the constraints from the PKS 2155-304 observation of \hess \cite{Abramowski:2013oea} and the NGC 1275 observation of \fermi \cite{TheFermi-LAT:2016zue} in Fig.~\ref{fig_contour}.
The combined limit from the Mrk 421 observations obtained here extends the excluded region to $g_{a\gamma} \gtrsim 3 \times 10^{-11}$ $\rm GeV^{-1}$ for $\sim 5\times 10^{-10}$ eV $\lesssim m_a \lesssim 2\times 10^{-9}$ eV and a part region for $m_a \simeq 2\times 10^{-7}$ eV below the CAST limit.
This constraint is also stricter than that derived from the observation data of PG 1553+113 \cite{Guo:2020kiq}.

\begin{figure}%%%%%%%%%%%%%%%%%%%%%%%%%%%%%%%%%%%%%%%%%FIG.bdtn1
\centering
\includegraphics[width=0.5\textwidth]{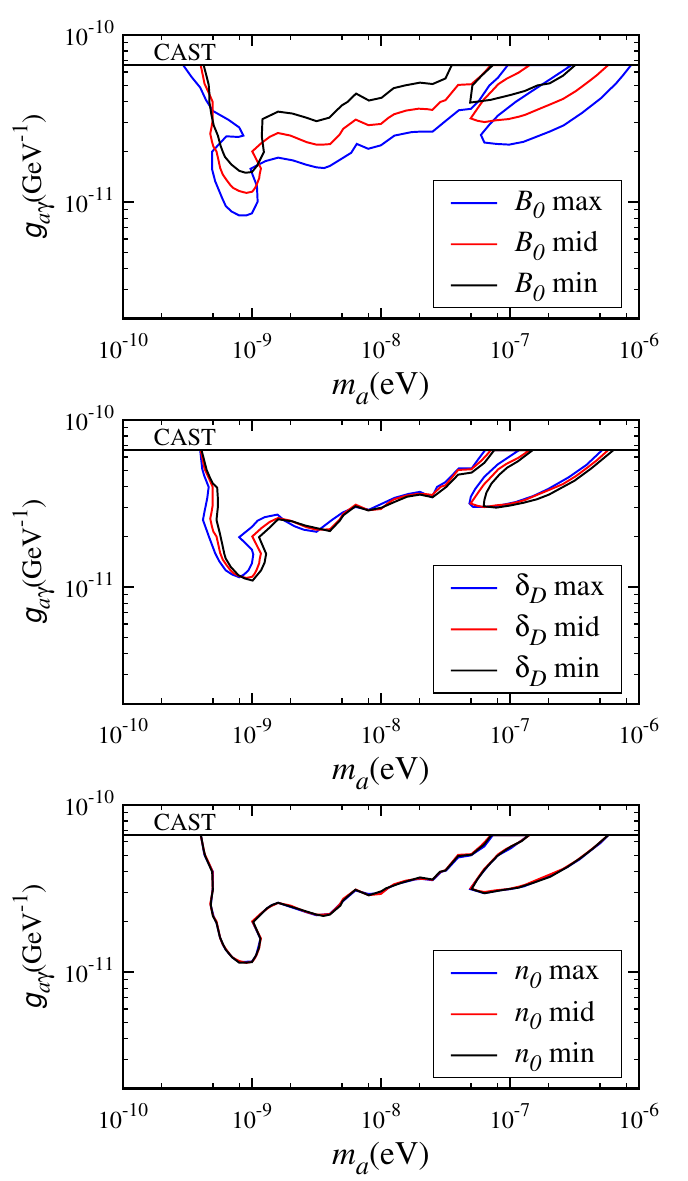}
\caption{The combined constraints at 95\% $\rm C.L.$ for the different values of $B_0$, $\delta_{\rm D}$, and $n_0$.
In each panel, we only change one parameter and take other parameters as the benchmark values listed in Table~\ref{tab_1}.
For the parameters $B_0$, $\delta_{\rm D}$, and $n_0$, the $1\sigma$ values in the minimal and maximal cases are given by Table~\ref{tab_4}.
The ``mid" lines represent the result with the benchmark values of all parameters listed in Table~\ref{tab_1}.
}
\label{fig_bdtn_1}
\end{figure}

\begin{figure}%%%%%%%%%%%%%%%%%%%%%%%%%%%%%%%%%%%%%%%%%FIG.bdtn2
\centering
\includegraphics[width=0.5\textwidth]{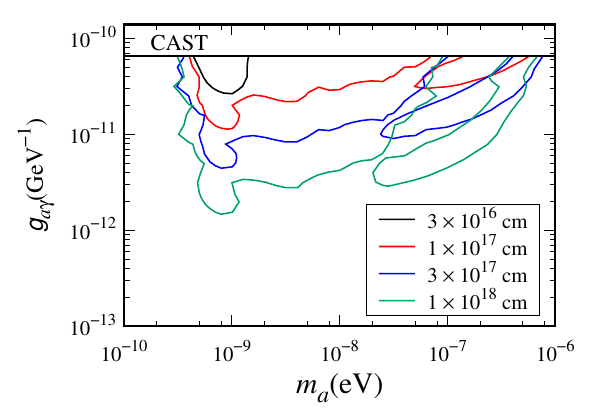}
\caption{The combined constraints at 95\% $\rm C.L.$ for the different values of $r_{\rm VHE}$. The lines from top to bottom represent the results for four typical values of $r_{\rm VHE} = 3\times 10^{16}$, $1\times 10^{17}$, $3\times 10^{17}$, and $1\times 10^{18}$ cm.
The other parameters for the BJMF model are taken as the benchmark values listed in Table~\ref{tab_1}.
}
\label{fig_bdtn_2}
\end{figure}

Finally, we discuss the impact of the uncertainties of the BJMF parameters.
The combined constraints on the ALP parameter space for different values of the BJMF parameters $B_0$, $\delta_{\rm D}$, $n_0$, and $r_{\rm VHE}$ are calculated.
In each panel of Fig.~\ref{fig_bdtn_1}, we only change the values of one parameter as listed in Table~\ref{tab_4} and take the other parameters as the benchmark values.
For the minimal (maximal) cases of $B_0$ and $\delta_{\rm D}$ with the uncertainties given by Ref.~\cite{Bartoli:2015cvo}, we take the $1\sigma$ minimal (maximal) values of the parameters in all the ten phases and calculate the corresponding constraints.
For the minimal (maximal) case of $n_0$, the values of $n_0$ listed in Table~\ref{tab_4} are calculated using the $1\sigma$ minimal (maximal) values of $u_e$ and $p_1$ given by Ref.~\cite{Bartoli:2015cvo}.
It can be seen that the constraints would become stringent for large $B_0$, while the changes of the other parameters do not significantly affect the results.

In Fig.~\ref{fig_bdtn_2}, we also show the impact of $r_{\rm VHE}$ on the final result.
Since there is no direct way to derive the precise value of $r_{\rm VHE}$, the uncertainty of this value would be larger than other parameters.
Here we choose other three typical values of $3\times 10^{16}$, $3\times 10^{17}$, and $1\times 10^{18}$ cm for $r_{\rm VHE}$. Since $r_{\rm VHE}$ directly characterizes the magnetic field strength and could significantly affects the ALP-photon oscillation effect, it can be seen that the large uncertainties of $r_{\rm VHE}$ significantly affect the final constraints.

%%%%%%%%%%%%%%%%%%%%%%%%%%%%%%%%%%%%%%%%%%%%%%%%%%%%%%TAB.4
\begin{table}[htb]
\caption{The modified values of the BJMF parameters in the ten phases used for Fig.~\ref{fig_bdtn_1}.}
\begin{ruledtabular}
\begin{tabular}{lllrlrl}
Phase &\multicolumn{2}{c}{$B_0$(G)}   &\multicolumn{2}{c}{ $ \delta_{\rm D} $}    & \multicolumn{2}{c}{$n_0$($\rm 10^{3} \ cm^{-3}$) }    \\
\hline
&min&max&min&max&min&max\\
\hline
S1   & 0.036  & 0.060  &  34    &  44   & 1.703   & 2.223 \\
S2   & 0.12   & 0.24   &  13    &  19   & 2.404   & 3.878 \\
OB   & 0.049  & 0.080  &  30    &  38   & 1.933   & 2.704 \\
F1   & 0.10   & 0.21   &  8     &  12   & 0.013   & 3.381 \\
F2   & 0.068  & 0.120  &  15    &  20   & 0.137   & 3.123 \\
F3   & 0.063  & 0.091  &  38    &  46   & 1.092   & 1.939 \\
F4   & 0.020  & 0.052  &  28    &  45   & 2.070   & 5.167 \\
F5   & 0.025  & 0.180  &  18    &  52   & 1.024   & 4.728 \\
F6   & 0.052  & 0.138  &  10    &  39   & 1.109   & 20.020 \\
F7   & 0.083  & 0.153  &  25    &  37   & 1.326   & 3.077 \\
\end{tabular}
\end{ruledtabular}
\label{tab_4}
\end{table}
%%%%%%%%%%%%%%%%%%%%%%%%%%%%%%%%%%%%%%%%%%%%%%%%%%%%%%%%%

\section{Conclusion}%%%%%%%%%%%%%%%%%%%%%%%%%%%%%%%%%%%%%%%%%Summary
\label{section_sum}

In this work, we investigate the ALP-photon oscillation effect in the spectra of the blazar Mrk 421 measured by \argo and \fermi during the common operation time, which cover ten activity phases of Mrk 421. We find that no significant ALP-photon oscillation effect is confirmed. However, only the observations of several phases can be individually used to set the constraint at 95\% $\rm C.L.$ on the ALP parameter space.

The constraint on the ALP-photon coupling $g_{a\gamma}$ becomes stricter when the analyses for the data of the ten phases are combined.
Compared with the limits set by the PKS 2155-304 observation of \hess \cite{Abramowski:2013oea} and the NGC 1275 observation of \fermi \cite{TheFermi-LAT:2016zue}, the upper limit on $g_{a\gamma}$ set by the \argo and \fermi observations of Mrk 421 is within $\sim [2\times 10^{-11}, \ 6\times 10^{-11}] \ \rm GeV^{-1}$ for the ALP mass of $\sim 5\times 10^{-10}$ eV $\lesssim m_a \lesssim 5\times 10^{-7}$ eV at 95\% $\rm C.L.$

We also show the impact of the BJMF parameters on the final constraints. 
We find that the final constraints are significantly affected by the 
magnetic field strength in the emission region $B_0$ and especially the distance of the  emission region to the central black hole $r_{\rm VHE}$.
The constraints would become more accurate when further information about these parameters will be available.

In the future, the forthcoming VHE $\gamma$-ray observations, such as CTA \cite{Acharya:2013sxa} and LHAASO \cite{Cao:2010zz} will collect more data for the high energy $\gamma$-ray sources at large distances from the Earth with high precision.
Using these results, it is possible to set the more stringent constraints on the ALP parameter space.

\section*{Acknowledgments}%%%%%%%%%%%%%%%%%%%%%%%%%%%%%%%%%%%%%%%%%Acknowledgments
The authors would like to thank Songzhan Chen for providing the energy spectra of Mrk 421 by \argo and \fermi in the common operation time.
This work is supported by the National Key R\&D Program of China (Grant No.~2016YFA0400200) and the National Natural Science Foundation of China (Grants No.~U1738209 and No.~11851303).

\bibliography{mrk_421}
\end{document}